# Twitter DM Videos Are Accessible to Unauthenticated Users

Michael L. Nelson
Old Dominion University
Norfolk VA 23529
mln@cs.odu.edu
@phonedude_mln

## Abstract

Videos shared in Twitter Direct Messages (DMs) have opaque URLs based on hashes of their content, but are otherwise available to unauthenticated HTTP users. These DM video URLs are thus hard to guess, but if they were somehow discovered, they are available to any user, including users without Twitter credentials (i.e., twitter.com specific HTTP Cookie or Authorization request headers). This includes web archives, such as the well-known Internet Archive Wayback Machine, which can be used to move DM videos to domains outside of twitter.com. This lack of authentication for DM videos is in contrast to Twitter's model for images in DMs, which also have opaque URLs but require a session-specific HTTP cookie shared only between the DM participants. We review a minimal reproducible example of an image and video shared between two demo accounts, and show that while the image is protected from unauthenticated access as well as from an authenticated third party, the video itself is persistently available for any user who knows the URL.

## Introduction

We have discovered that although Twitter direct messages (DMs) have opaque URLs that would make them difficult to guess, the videos shared in DMs are not protected by any form of HTTP authentication, requiring neither twitter.com specific Cookie or Authorization HTTP request headers.[1] This is in contrast to how Twitter uses session-specific HTTP Cookie values to protect images, even though their URLs are also named using hashes and are thus presumably similarly difficult to guess or reverse engineer. In effect, while images in DMs are highly protected from third party access, requiring the user to be a party to the DMs *as well as requiring the DM session to be currently active,* videos in DMs are protected only through "security through obscurity".

[1] The initial version of this document was developed as a Google Doc, with a .pdf version exported on 2022-12-09. The initial .pdf document has its hash posted in a reply to https://twitter.com/phonedude_mln/status/1601268082424487936. The 2022-12-10 version was updated to reflect Twitter's response (Appendix 3), closing report #1798935, and was submitted to arXiv.org. This version includes a review of HTTP (i.e., not HTTPS) support on video.twimg.com and a discussion of Twitter DM videos already archived at the Internet Archive.

Below, we introduce a minimal reproducible example using three demo accounts, two of which exchange via DMs text, an image, and a video. We show that the image is available only to authenticated parties while the DM chat session is open. We show that the video is available to any party, regardless of authentication and session activity. We retrieve the video using curl commands and reassemble the video shared in the DM chat session. We then push the video URLs into the Internet Archive's Wayback Machine, using their Save Page Now service [2]. We retrieve the archived video fragments, reassemble the video, and show that it is byte-for-byte equivalent to the live web version. The implication is that until video URLs are protected in the same way as image URLs, they are vulnerable to exfiltration to non-twitter.com machines without the knowledge of or participation by the DM participants.

While it seems unlikely that Twitter video URLs could be guessed, it is possible that the URLs themselves could leak via some heretofore unknown mechanism, vulnerability, or social engineering. It is also possible that the URL naming mechanism could be reverse engineered, subject to brute-force attacks, or some combination of the above. Once the URLs are known, they are not protected.

# A Twitter DM Session

We created three Twitter accounts, each corresponding to one of our cats (Appendix 1). These accounts were created solely for this demonstration, and may be deleted in the future. The accounts that are sending DMs are [@WhiteKitty2012](#) and [@BKitty2020](#), with [@kitten2017_](#) later attempting to access the resources shared in those DMs.

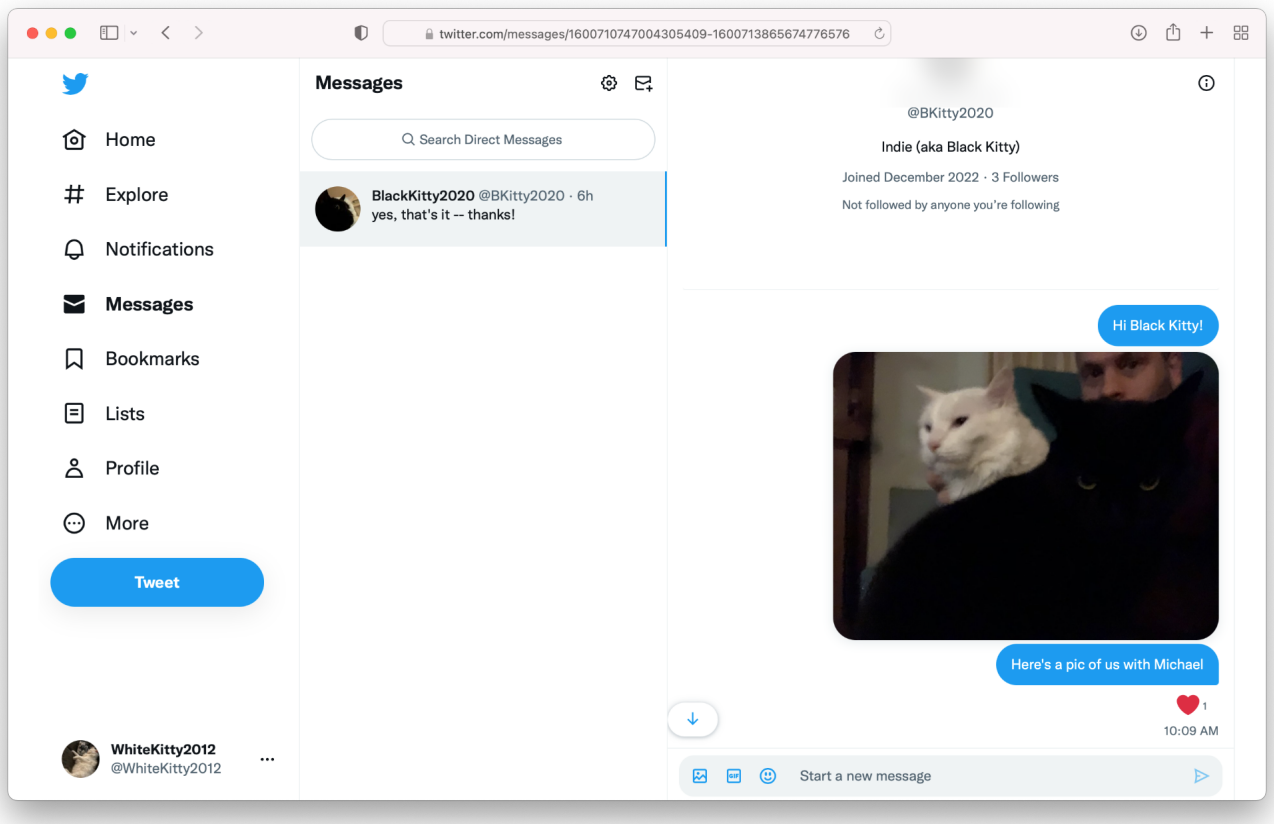

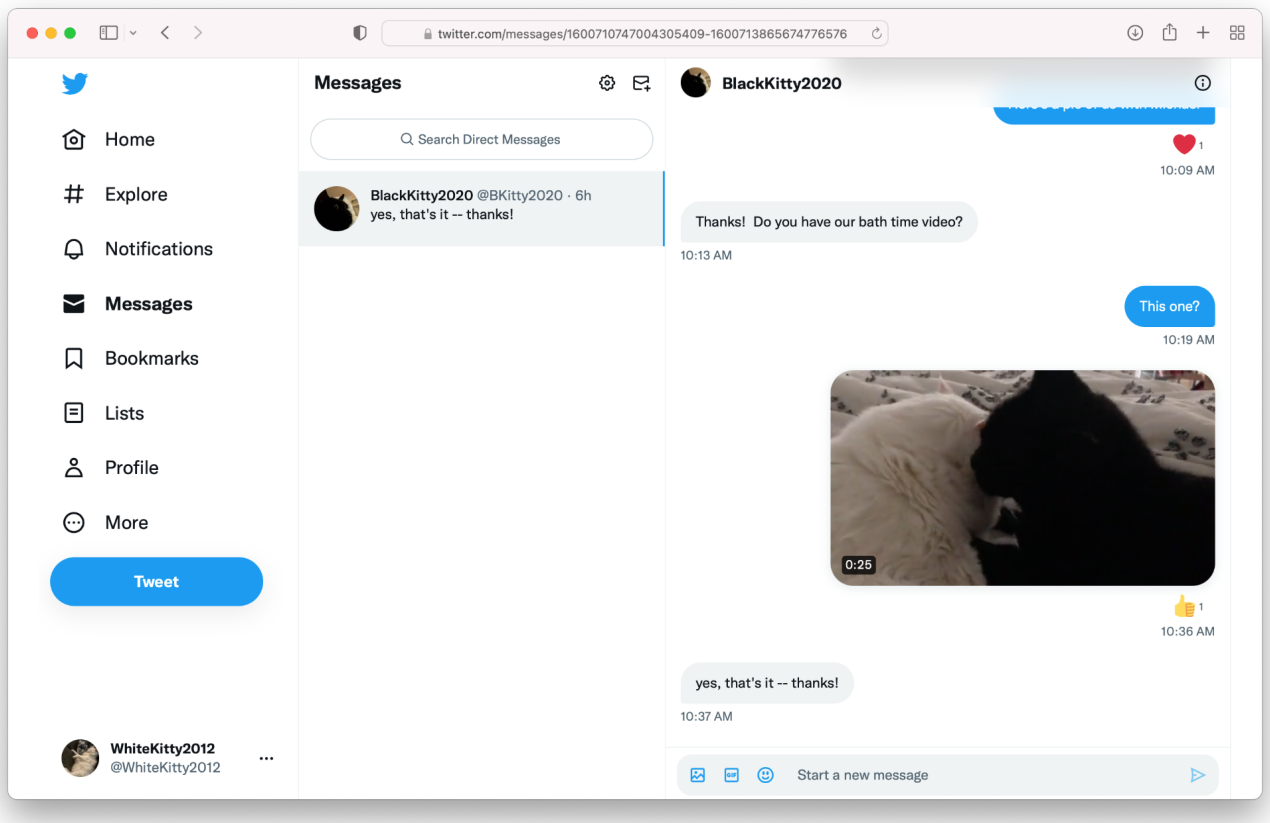

Figure 1: DMs between @WhiteKitty2012 and @BKitty2020; the top shows a JPEG sent by @WhiteKitty2012 and the bottom shows a 32 second video sent by @WhiteKitty2012.

Figure 1 shows the initial exchange between the two accounts. @WhiteKitty2012 sends, in separate DMs, a `.jpeg` and `.mov` file (`.mov` is converted to `.mp4` at twitter.com).

# The Image in the DMs

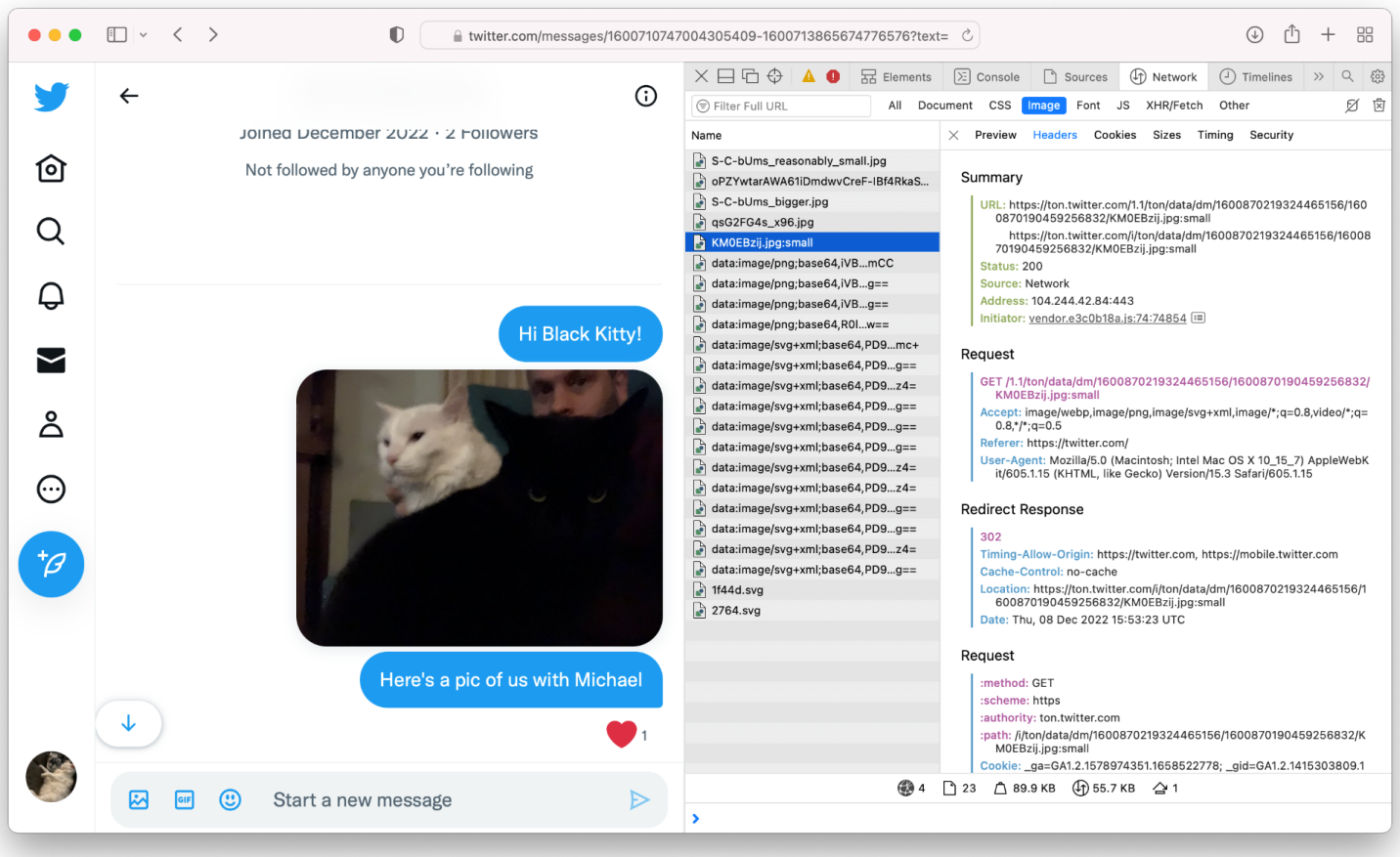

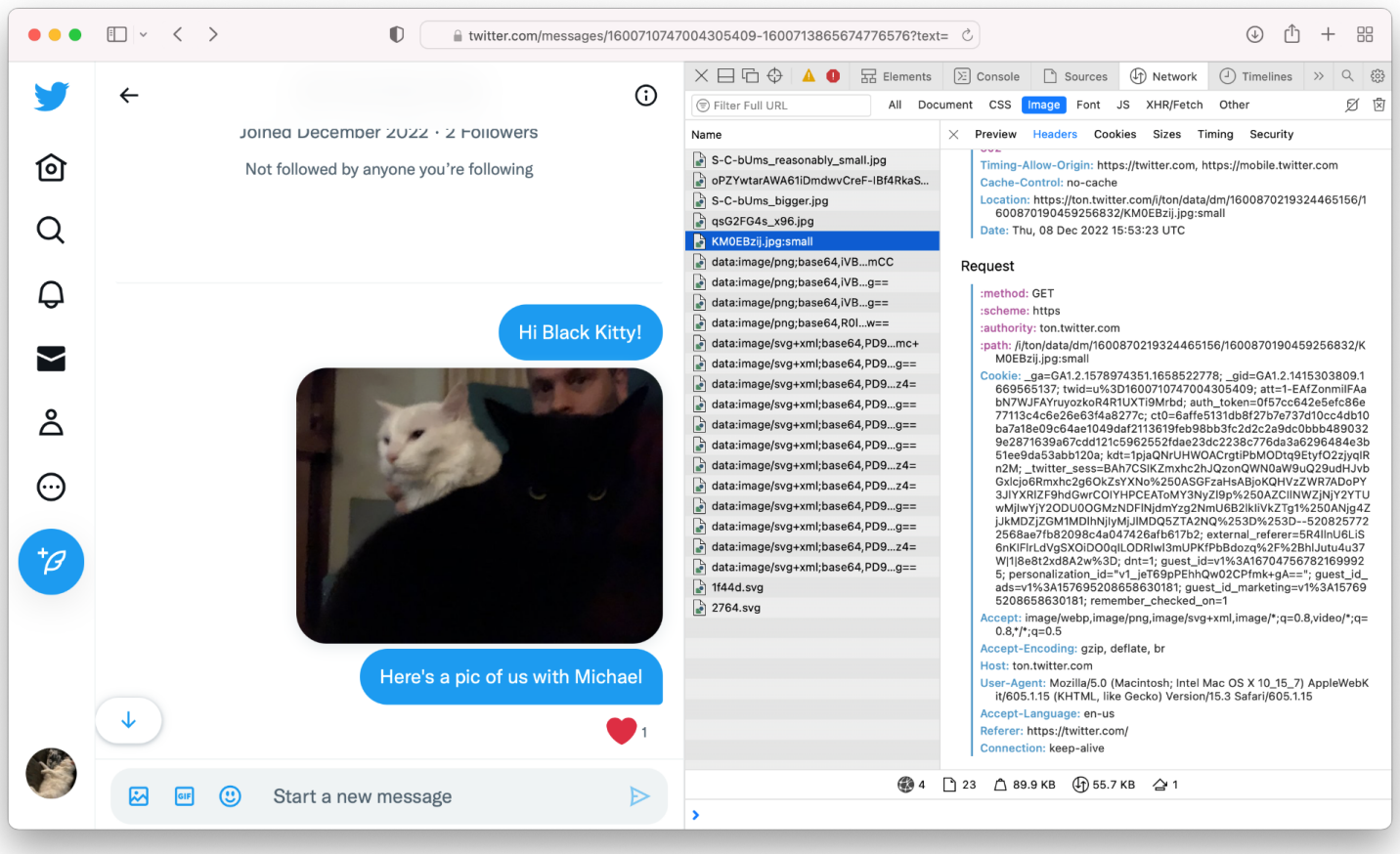

Figure 2: Examining the resources revealed via the developer tools. The lower image is the same, just scrolled down.

Figure 2 shows the DM exchange with the Safari developer tools open. From the tools, we can see that the image URL is:

```
https://ton.twitter.com/i/ton/data/dm/1600870219324465156/1600870190459256832/KM0EBzij.jpg:small
```

If we try to curl that URL without cookies set from twitter.com, we will get an HTTP 401 Unauthorized response:

```
% curl -iL
https://ton.twitter.com/i/ton/data/dm/1600870219324465156/1600870190459256832/KM0EBzij.jpg:small
HTTP/2 401
date: Thu, 08 Dec 2022 15:59:47 UTC
perf: 7626143928
server: tsa_b
set-cookie: guest_id_marketing=v1%3A167051518741457080; Max-Age=63072000; Expires=Sat, 07 Dec 2024
15:59:47 GMT; Path=/; Domain=.twitter.com; Secure; SameSite=None
set-cookie: guest_id_ads=v1%3A167051518741457080; Max-Age=63072000; Expires=Sat, 07 Dec 2024 15:59:47
GMT; Path=/; Domain=.twitter.com; Secure; SameSite=None
set-cookie: personalization_id="v1_hwtdvArizEra8KCDu8hSrg=="; Max-Age=63072000; Expires=Sat, 07 Dec
2024 15:59:47 GMT; Path=/; Domain=.twitter.com; Secure; SameSite=None
set-cookie: guest_id=v1%3A167051518741457080; Max-Age=63072000; Expires=Sat, 07 Dec 2024 15:59:47 GMT;
Path=/; Domain=.twitter.com; Secure; SameSite=None
cache-control: no-cache
content-length: 0
x-transaction-id: f7ba97afd6681c13
timing-allow-origin: https://twitter.com, https://mobile.twitter.com
x-content-type-options: nosniff
strict-transport-security: max-age=631138519
x-response-time: 6
x-connection-hash: 15a2dec13205b54f59fff1c45934e28f0477447d4d2851b1005452dfde31b672
```

Using the "copy as curl" option in the developer tools, we can access this image because it includes the necessary HTTP cookies. In the command below, the output has been redirected to the file "wk-bk.jpeg", and the ASCII progress bar from curl shows that 22425 bytes have been received.

```
% curl
'https://ton.twitter.com/i/ton/data/dm/1600870219324465156/1600870190459256832/KM0EBzij.jpg:small' \
-X 'GET' \
-H 'Cookie: _ga=GA1.2.1578974351.1658522778; _gid=GA1.2.1415303809.1669565137;
twid=u%3D1600710747004305409; att=1-EAfZonmiIFAabN7
WJFAYruyozkoR4R1UXTi9Mrbd; auth_token=0f57cc642e5efc86e77113c4c6e26e63f4a8277c;
ct0=6affe5131db8f27b7e737d10cc4db10ba7a18e09c64ae1
049daf2113619feb98bb3fc2d2c2a9dc0bbb4890329e2871639a67cdd121c5962552fdae23dc2238c776da3a6296484e3b51ee9
da53abb120a; kdt=1pjaQNrUHW
OACrgtiPbMODtq9EtyfO2zjyqIRn2M;
_twitter_sess=BAh7CSIKZmxhc2hJQzonQWN0aW9uQ29udHJvbGxlcjo6Rmxhc2g6OkZsYXNo%250ASGFzaHsABjoKQHVzZWR
7ADoPY3JlYXRlZF9hdGwrCOIYHPCEAToMY3NyZl9p%250AZCIlNWZjNjY2YTUwMjIwYjY2ODU0OGMzNDFlNjdmYzg2NmU6B2lkIiVkZ
Tg1%250ANjg4ZjJkMDZjZGM1MDl
hNjIyMjJlMDQ5ZTA2NQ%253D%253D--5208257722568ae7fb82098c4a047426afb617b2;
external_referer=5R4llnU6LiS6nKIFlrLdVgSXOiDO0qILODRlwl3m
UPKfPbBdozq%2F%2BhlJutu4u37W|1|8e8t2xd8A2w%3D; dnt=1; guest_id=v1%3A167047567821699925;
personalization_id="v1_jeT69pPEhhQw02CPfmk
+gA=="; guest_id_ads=v1%3A157695208658630181; guest_id_marketing=v1%3A157695208658630181;
remember_checked_on=1' \
-H 'Accept: image/webp,image/png,image/svg+xml,image/*;q=0.8,video/*;q=0.8,*/*;q=0.5' \
-H 'Accept-Encoding: gzip, deflate, br' \
-H 'Host: ton.twitter.com' \
-H 'User-Agent: Mozilla/5.0 (Macintosh; Intel Mac OS X 10_15_7) AppleWebKit/605.1.15 (KHTML, like
Gecko) Version/15.3 Safari/605.1.15' \
-H 'Accept-Language: en-us' \
-H 'Referer: https://twitter.com/' \
-H 'Connection: keep-alive' > wk-bk.jpeg
  % Total    % Received % Xferd  Average Speed   Time    Time     Time  Current
                                 Dload  Upload   Total   Spent    Left  Speed
100 22425  100 22425    0     0   109k      0 --:--:-- --:--:-- --:--:--  109k
% open wk-bk.jpeg
```

Some web servers are designed to refuse robot access, and if the request does not look like it comes from an interactive browser (e.g., Chrome, Firefox, Mozilla) [3], the server will send an HTTP 401 response or some other non-200 OK response. In the example below, we remove the Cookie request header, but keep all the other request headers that indicate an interactive user (e.g., User-Agent, Accept, Referer (sic)), but we still receive a 401 response:

```
% curl -i
'https://ton.twitter.com/i/ton/data/dm/1600870219324465156/1600870190459256832/KM0EBzij.jpg:small' \
-X 'GET' \
-H 'Accept: image/webp,image/png,image/svg+xml,image/*;q=0.8,video/*;q=0.8,*/*;q=0.5' \
-H 'Accept-Encoding: gzip, deflate, br' \
-H 'Host: ton.twitter.com' \
-H 'User-Agent: Mozilla/5.0 (Macintosh; Intel Mac OS X 10_15_7) AppleWebKit/605.1.15 (KHTML, like
Gecko) Version/15.3 Safari/605.1.15' \
-H 'Accept-Language: en-us' \
-H 'Referer: https://twitter.com/' \
-H 'Connection: keep-alive'
HTTP/2 401
date: Thu, 08 Dec 2022 16:02:03 UTC
perf: 7626143928
server: tsa_b
set-cookie: guest_id_marketing=v1%3A167051532319121789; Max-Age=63072000; Expires=Sat, 07 Dec 2024
16:02:03 GMT; Path=/; Domain=.twitter.com; Secure; SameSite=None
set-cookie: guest_id_ads=v1%3A167051532319121789; Max-Age=63072000; Expires=Sat, 07 Dec 2024 16:02:03
GMT; Path=/; Domain=.twitter.com; Secure; SameSite=None
set-cookie: personalization_id="v1_h/6bss49bkzROqTp2DIipw=="; Max-Age=63072000; Expires=Sat, 07 Dec
2024 16:02:03 GMT; Path=/; Domain=.twitter.com; Secure; SameSite=None
set-cookie: guest_id=v1%3A167051532319121789; Max-Age=63072000; Expires=Sat, 07 Dec 2024 16:02:03 GMT;
Path=/; Domain=.twitter.com; Secure; SameSite=None
cache-control: no-cache
content-length: 0
```

```
x-transaction-id: 92d4a204724483c2
timing-allow-origin: https://twitter.com, https://mobile.twitter.com
x-content-type-options: nosniff
strict-transport-security: max-age=631138519
x-response-time: 5
x-connection-hash: 7eec83c817183d85f2f9b8bd2a95b118e3c971c22d34777b004f812accacb42d
```

To reduce access from non-twitter.com sites, we can see removing the Referer request header will cause the server to return a 404 instead of a 401:

```
% curl -i
'https://ton.twitter.com/i/ton/data/dm/1600870219324465156/1600870190459256832/KM0EBzij.jpg:small' \
-X 'GET' \
-H 'Cookie: _ga=GA1.2.1578974351.1658522778; _gid=GA1.2.1415303809.1669565137;
twid=u%3D1600710747004305409; att=1-kHxG28wDTAl7dHp4NwjtSfKxlx3iHoQeoQbs2o5z;
auth_token=637657abf4a99fb7ee02b978448244f04e4d4a00;
ct0=a958f95b304a239facb5efb101533415223e51041e2b2b14259d3821dc81c9227a0568c3821826e9726f1cf8b9cfae1289a
e5993034988f53d6d6a5a15eded475f90cafb9f63f979c6255b7ed90c9369;
kdt=1pjaQNrUHWOACrgtiPbMODtq9EtyfO2zjyqIRn2M;
_twitter_sess=BAh7CSIKZmxhc2hJQzonQWN0aW9uQ29udHJvbGxlcjo6Rmxhc2g6OkZsYXNo%250ASGFzaHsABjoKQHVzZWR7ADoP
Y3JlYXRlZF9hdGwrCI0qxfOEAToMY3NyZl9p%250AZCIlMjk0YTY5ZDJlNGM0YmEzZWFkYTRiMjBkZmNhNjZkOGM6B2lkIiUyM2Yw%2
50AMDlhMGQxYjVmNWNmMzczYmRmMjc3YmRjOWU4MQ%253D%253D--503caeb7e058e15aadc86f2267123435bd7e2b1d;
gt=1600974721537003538; dnt=1; guest_id=v1%3A167053709193512452;
personalization_id="v1_urbaNOacunAsjDYcF7mUyg=="; guest_id_ads=v1%3A157695208658630181;
guest_id_marketing=v1%3A157695208658630181; remember_checked_on=1' \
-H 'Accept: image/webp,image/png,image/svg+xml,image/*;q=0.8,video/*;q=0.8,*/*;q=0.5' \
-H 'Host: ton.twitter.com' \
-H 'User-Agent: Mozilla/5.0 (Macintosh; Intel Mac OS X 10_15_7) AppleWebKit/605.1.15 (KHTML, like
Gecko) Version/15.3 Safari/605.1.15' \
-H 'Accept-Language: en-us' \
-H 'Connection: keep-alive'
HTTP/2 404
date: Thu, 08 Dec 2022 23:36:52 UTC
perf: 7626143928
server: tsa_b
cache-control: no-cache
content-length: 0
x-transaction-id: 01e6757f5999c4a0
timing-allow-origin: https://twitter.com, https://mobile.twitter.com
x-content-type-options: nosniff
strict-transport-security: max-age=631138519
x-response-time: 8
x-connection-hash: 16948b5c906bb29a772a3585443f921d2a9b7a915b8b99674d51f36e6e946d22
```

This is also shown in Figure 3, where a simple HTML file that attempts to include the image from a non-twitter.com host:

```
% more grab-dm-image.html
hello world

<img
src="https://ton.twitter.com/i/ton/data/dm/1600870219324465156/1600870190459256832/KM0EBzij.jpg:small">
```

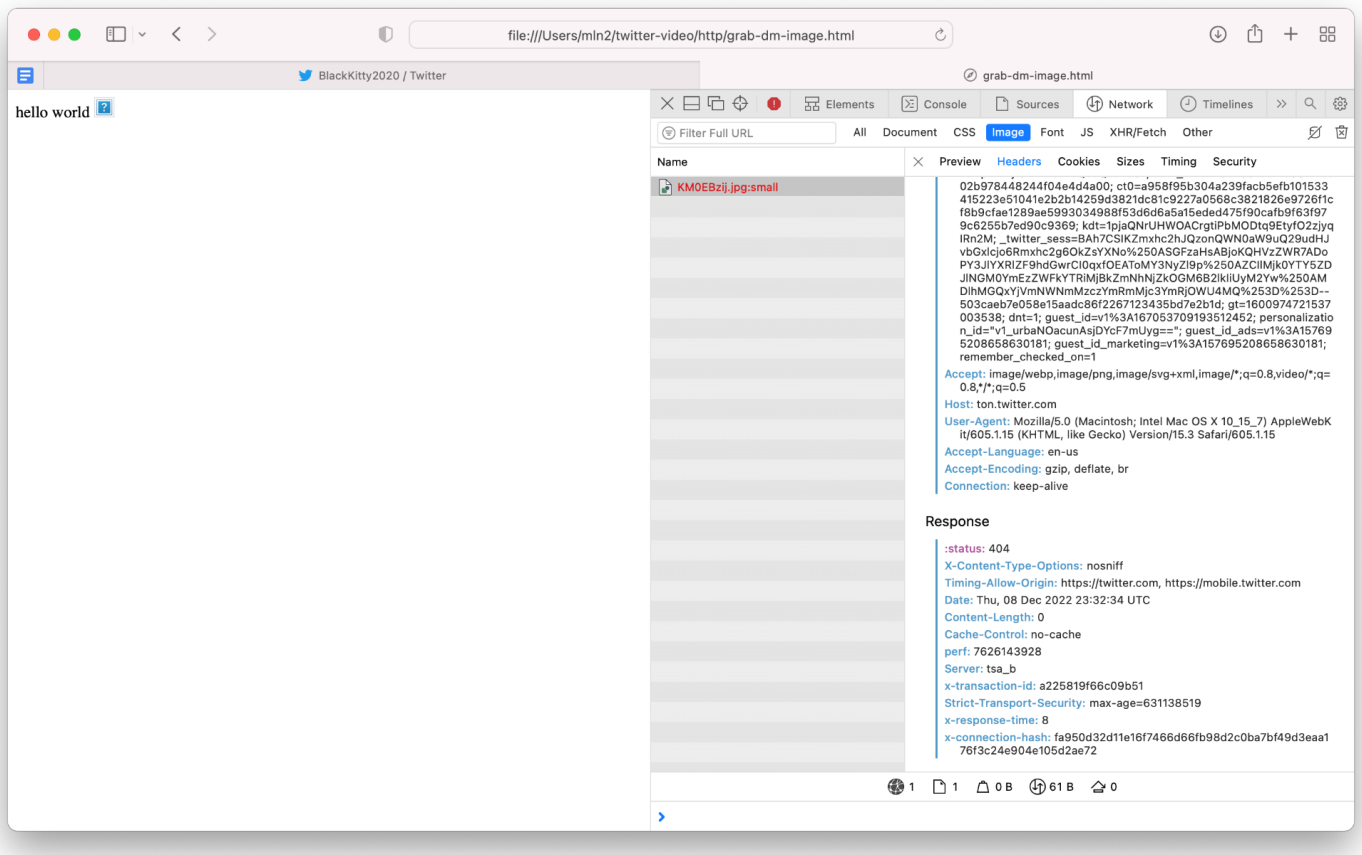

Figure 3: A 404 response is returned when the request does not have the correct Referer request header.

Repeating the successful curl request from above, if the session is recently closed, the server will return a 401:

```
% curl -i
'https://ton.twitter.com/i/ton/data/dm/1600870219324465156/1600870190459256832/KM0EBzij.jpg:small' \
-X 'GET' \
-H 'Cookie: _ga=GA1.2.1578974351.1658522778; _gid=GA1.2.1415303809.1669565137;
twid=u%3D1600710747004305409; att=1-n11EVoxa8oM767165HUbjPZ8YJFiYvFZMzw3uCFr; auth_token=88ba7e40f9067e
ec06c5501f86da52c9658dbf3b;
ct0=43a7ed19a7f2dd7d0f8cf464bb9ac2b31fcacf0288a38c2c8a0773a2327a4f7d82b02030b5255758b42c5cd26ff817784a4
88a3fbd108dff3963f4ba66b3a1d46cd2ac8031028e57f6a3bd
25253aae40; kdt=1pjaQNrUHWOACrgtiPbMODtq9EtyfO2zjyqIRn2M;
_twitter_sess=BAh7CSIKZmxhc2hJQzonQWN0aW9uQ29udHJvbGxlcjo6Rmxhc2g6OkZsYXNo%250ASGFzaHsABjoKQHVzZWR7ADoP
Y3JlYXRlZF9hdGwrCEj9I
POEAToMY3NyZl9p%250AZCIlZWQ5OGM3NWE3NjliNWM5YTExYTAyODEwZjEzMzE2ZmQ6B2lkIiUxOGI2%250AZjk4Y2Q2NTlmMjI1YW
QxNDAxZTEyNjhhNjI5NA%253D%253D--8432d160885b961f50dd7ad4bfac669c10cd8f1b; gt=16
00929594542067725; dnt=1; guest_id=v1%3A167052633266899352;
personalization_id="v1_I8bVcj9tTkUu6IElXKvpxA=="; guest_id_ads=v1%3A157695208658630181;
guest_id_marketing=v1%3A1576952086
58630181; remember_checked_on=1' \
-H 'Accept: image/webp,image/png,image/svg+xml,image/*;q=0.8,video/*;q=0.8,*/*;q=0.5' \
-H 'Host: ton.twitter.com' \
-H 'User-Agent: Mozilla/5.0 (Macintosh; Intel Mac OS X 10_15_7) AppleWebKit/605.1.15 (KHTML, like
Gecko) Version/15.3 Safari/605.1.15' \
-H 'Accept-Language: en-us' \
-H 'Referer: https://twitter.com/' \
-H 'Connection: keep-alive'
HTTP/2 401
date: Thu, 08 Dec 2022 19:20:26 UTC
perf: 7626143928
server: tsa_b
cache-control: no-cache
content-length: 0
x-transaction-id: b7637e1d0cca1353
timing-allow-origin: https://twitter.com, https://mobile.twitter.com
x-content-type-options: nosniff
strict-transport-security: max-age=631138519
x-response-time: 5
x-connection-hash: 5161af05a9e051d7c2df362290b685cdbe9c1722ef4c0b836b96a54070d4fa0f
```

And after more time has passed, the same request will not even receive an HTTP response:

```
% curl -i
'https://ton.twitter.com/i/ton/data/dm/1600870219324465156/1600870190459256832/KM0EBzij.jpg:small' \
-X 'GET' \
-H 'Cookie: _ga=GA1.2.1578974351.1658522778; _gid=GA1.2.1415303809.1669565137;
twid=u%3D1600710747004305409; att=1-n11EVoxa8oM767165HUbjPZ8YJFiYvFZMzw3uCFr; auth_token=88ba7e40f9067e
ec06c5501f86da52c9658dbf3b;
ct0=43a7ed19a7f2dd7d0f8cf464bb9ac2b31fcacf0288a38c2c8a0773a2327a4f7d82b02030b5255758b42c5cd26ff817784a4
88a3fbd108dff3963f4ba66b3a1d46cd2ac8031028e57f6a3bd
25253aae40; kdt=1pjaQNrUHWOACrgtiPbMODtq9EtyfO2zjyqIRn2M;
_twitter_sess=BAh7CSIKZmxhc2hJQzonQWN0aW9uQ29udHJvbGxlcjo6Rmxhc2g6OkZsYXNo%250ASGFzaHsABjoKQHVzZWR7ADoP
Y3JlYXRlZF9hdGwrCEj9I
POEAToMY3NyZl9p%250AZCIlZWQ5OGM3NWE3NjliNWM5YTExYTAyODEwZjEzMzE2ZmQ6B2lkIiUxOGI2%250AZjk4Y2Q2NTlmMjI1YW
QxNDAxZTEyNjhhNjI5NA%253D%253D--8432d160885b961f50dd7ad4bfac669c10cd8f1b; gt=16
00929594542067725; dnt=1; guest_id=v1%3A167052633266899352;
personalization_id="v1_I8bVcj9tTkUu6IElXKvpxA=="; guest_id_ads=v1%3A157695208658630181;
guest_id_marketing=v1%3A1576952086
58630181; remember_checked_on=1' \
-H 'Accept: image/webp,image/png,image/svg+xml,image/*;q=0.8,video/*;q=0.8,*/*;q=0.5' \
-H 'Host: ton.twitter.com' \
-H 'User-Agent: Mozilla/5.0 (Macintosh; Intel Mac OS X 10_15_7) AppleWebKit/605.1.15 (KHTML, like
Gecko) Version/15.3 Safari/605.1.15' \
-H 'Accept-Language: en-us' \
-H 'Referer: https://twitter.com/' \
-H 'Connection: keep-alive'

curl: (52) Empty reply from server
```

Finally, while the session is active between @WhiteKitty2012 and @BKitty2020, below we substitute a Cookie request header borrowed from a session from @Kitten2017_. The server returns a 404 response instead of a 401, indicating that while there is valid authentication to twitter.com (via the cookie from @Kitten2017_), the resource is not available to @Kitten2017_.

```
% curl -i
'https://ton.twitter.com/i/ton/data/dm/1600870219324465156/1600870190459256832/KM0EBzij.jpg:small' \
-X 'GET' \
-H 'Cookie: guest_id_marketing=v1%3A166110673717544305; guest_id_ads=v1%3A166110673717544305;
_ga=GA1.2.246184186.1661106740; kdt=mGdhDHJWU2Xojs0WFHgTqInVyjmtyyEnz2HjdiSm;
mbox=PC#7d0f03dfc0b543c280a8d814638bf284.34_0#1733508101|session#0d375b037bfc4affa5317cec62d5088f#16702
65161; _ga_BYKEBDM7DS=GS1.1.1670263302.2.0.1670263307.0.0.0; dnt=1; _gid=GA1.2.1467843531.1668961873;
_ga_34PHSZMC42=GS1.1.1670273443.2.1.1670273444.0.0.0; at_check=true; lang=en;
personalization_id="v1_p3BDKhaBPoTfWxbxGeqWZA=="; guest_id=v1%3A167055513085263151;
ct0=c018e65cde223ea950ff5edd0064b2a0181a22d3d91e483e3a4c9a6191b1be9b35a64e5a1b268df5ae40226684cdac8a12e
0c966c89de6c1ed74142da72bdf9ea4a769d175291745024b668262ca78be; gt=1601050381722730496;
_twitter_sess=BAh7CSIKZmxhc2hJQzonQWN0aW9uQ29udHJvbGxlcjo6Rmxhc2g6OkZsYXNo%250ASGFzaHsABjoKQHVzZWR7ADoP
Y3JlYXRlZF9hdGwrCOO42PSEAToMY3NyZl9p%250AZCIlN2IzZDk5NDM3ZmRiNzc0ZWM4NGI0YzFlNWMyZmM4MDc6B2lkIiViODQw%2
50AMDM1NjJkNDgxNGIwNTM1ODQ1NmU5NzlmZTQwMg%253D%253D--616c7ec3658da39d70a1f4659d254950f00530fa;
twid=u%3D1600918912891293698; auth_token=40a7ccfccee035bf51d86fedc2035a6695286e0e;
att=1-ewEJN9TreAXQvwHuXJEwxpNTn2pYnWWq4y0W1ZHy' \
-H 'Accept: image/webp,image/png,image/svg+xml,image/*;q=0.8,video/*;q=0.8,*/*;q=0.5' \
-H 'Accept-Encoding: gzip, deflate, br' \
-H 'Host: ton.twitter.com' \
-H 'User-Agent: Mozilla/5.0 (Macintosh; Intel Mac OS X 10_15_7) AppleWebKit/605.1.15 (KHTML, like
Gecko) Version/15.3 Safari/605.1.15' \
-H 'Accept-Language: en-us' \
-H 'Referer: https://twitter.com/' \
-H 'Connection: keep-alive'
HTTP/2 404
date: Fri, 09 Dec 2022 03:28:50 UTC
perf: 7626143928
server: tsa_b
```

```
cache-control: no-cache
content-length: 0
x-transaction-id: db3608313b0eabae
timing-allow-origin: https://twitter.com, https://mobile.twitter.com
x-content-type-options: nosniff
strict-transport-security: max-age=631138519
x-response-time: 15
x-connection-hash: 92e4ac9909a9d9801798abcabd132169f18e19d895374160a9dbad499938f1fb
```

# The Video in the DMs

The previous section documents the extent to which twitter.com will protect images from unauthorized access from outside the DM session in which images are shared. We show that twitter.com does not afford the same protection to videos. In Figure 4, we show the developer tools revealing the browser fetching various video resources. The one we highlight is a Media Playlist for a fragmented MPEG-4 (as defined in RFC 8216 [4]).

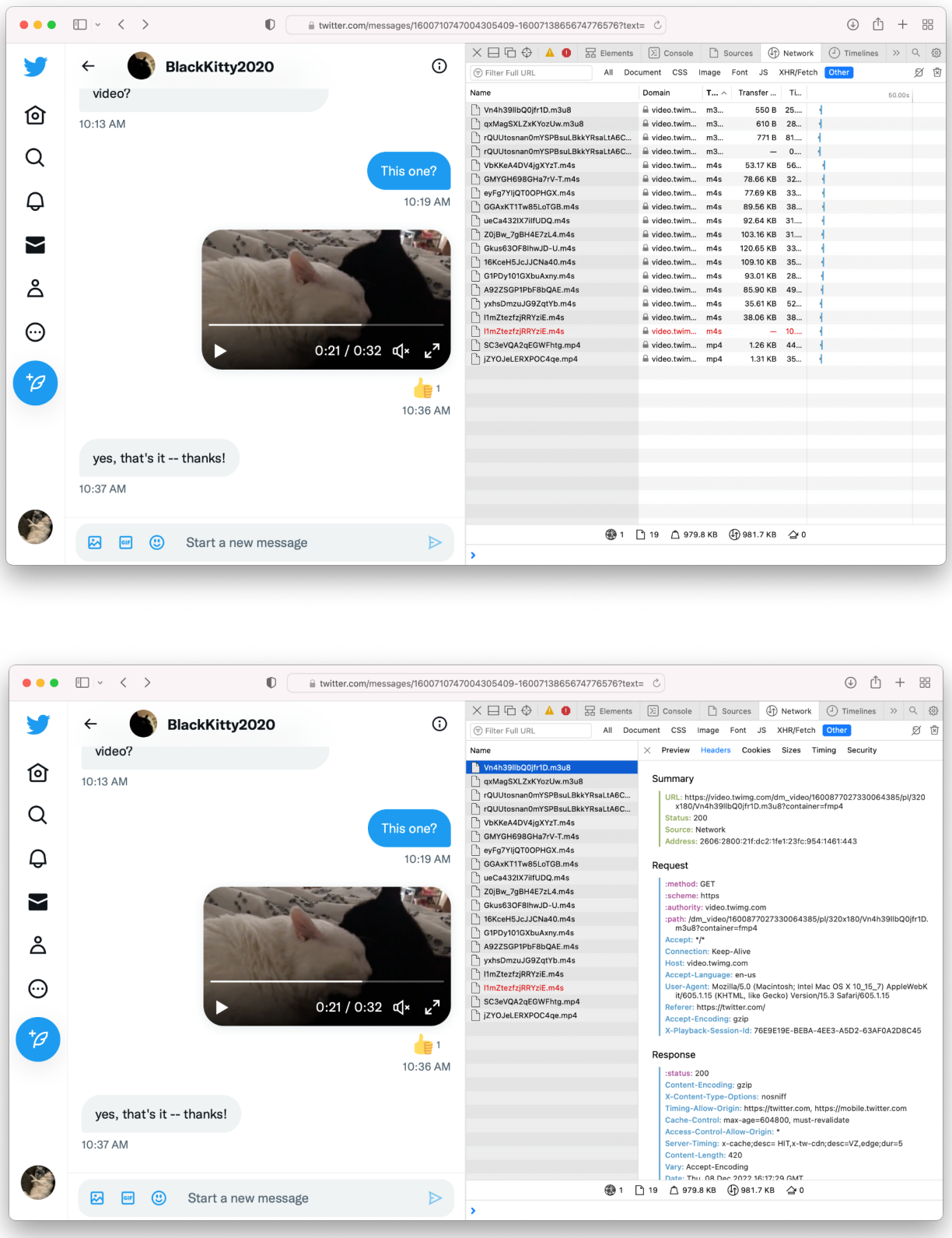

Figure 4: Developer tool view for video resources; the highlighted resource is a Media Playlist.

We can dereference the URL for the Media Playlist without authentication (i.e., with a simple curl command) and the Media Playlist contains the relative URLs for the fragmented MP4: the single Media Initialization Section (the `.mp4` file in the `EXT-X-MAP` field) and the multiple

Media Segments (the `.m4s files` in the `EXINF` field). The listing below shows a simple shell script that curls the .mp4 and .m4s files, without authentication, and saves them locally. They can then be concatenated and the original video is recovered and playable.

```
% curl -i
"https://video.twimg.com/dm_video/1600877027330064385/pl/320x180/Vn4h39llbQ0jfr1D.m3u8?container=fmp4"
HTTP/2 200
perf: 7626143928
content-type: application/x-mpegURL
cache-control: max-age=604800, must-revalidate
last-modified: Thu, 08 Dec 2022 15:34:40 GMT
x-transaction-id: 8d01a92de1228077
timing-allow-origin: https://twitter.com, https://mobile.twitter.com
x-content-type-options: nosniff
strict-transport-security: max-age=631138519
access-control-allow-origin: *
access-control-expose-headers: Content-Length
accept-ranges: bytes
date: Thu, 08 Dec 2022 16:21:09 GMT
x-served-by: cache-fty21349-FTY, cache-iad-kiad7000031-IAD
x-cache: MISS, MISS
vary: Accept-Encoding
x-tw-cdn: FT
server-timing: x-cache;desc=MISS, x-tw-cdn;desc=FT
content-length: 1177

#EXTM3U
#EXT-X-VERSION:6
#EXT-X-MEDIA-SEQUENCE:0
#EXT-X-TARGETDURATION:3
#EXT-X-PLAYLIST-TYPE:VOD
#EXT-X-MAP:URI="/dm_video/1600877027330064385/vid/0/0/320x180/jZYOJeLERXPOC4qe.mp4"
#EXTINF:3.000,
/dm_video/1600877027330064385/vid/0/3000/320x180/l1mZtezfzjRRYziE.m4s
#EXTINF:3.000,
/dm_video/1600877027330064385/vid/3000/6000/320x180/yxhsDmzuJG9ZqtYb.m4s
#EXTINF:3.000,
/dm_video/1600877027330064385/vid/6000/9000/320x180/KGtEzr2KaRfP4Y6H.m4s
#EXTINF:3.000,
/dm_video/1600877027330064385/vid/9000/12000/320x180/V1cXuDCxjXKk_JT9.m4s
#EXTINF:3.000,
/dm_video/1600877027330064385/vid/12000/15000/320x180/CtUTfpf83EHjEFjd.m4s
#EXTINF:3.000,
/dm_video/1600877027330064385/vid/15000/18000/320x180/lGncur15MHC6fvKg.m4s
#EXTINF:3.000,
/dm_video/1600877027330064385/vid/18000/21000/320x180/2U51PtuljYpAxhRr.m4s
#EXTINF:3.000,
/dm_video/1600877027330064385/vid/21000/24000/320x180/gs-8pNvThX_1kjx0.m4s
#EXTINF:3.000,
/dm_video/1600877027330064385/vid/24000/27000/320x180/f1UDV6NW3odTL-ux.m4s
#EXTINF:3.000,
/dm_video/1600877027330064385/vid/27000/30000/320x180/9hFdPsP3QYV8c130.m4s
#EXTINF:2.100,
/dm_video/1600877027330064385/vid/30000/32100/320x180/yFOkibojJs9PWhkX.m4s
#EXT-X-ENDLIST
```

```
% cat run-me.sh
#!/bin/csh -x

curl "https://video.twimg.com/dm_video/1600877027330064385/vid/0/0/320x180/jZYOJeLERXPOC4qe.mp4" >
00.mp4
curl "https://video.twimg.com/dm_video/1600877027330064385/vid/0/3000/320x180/l1mZtezfzjRRYziE.m4s " >
01.m4s
curl "https://video.twimg.com/dm_video/1600877027330064385/vid/3000/6000/320x180/yxhsDmzuJG9ZqtYb.m4s "
> 02.m4s
curl "https://video.twimg.com/dm_video/1600877027330064385/vid/6000/9000/320x180/KGtEzr2KaRfP4Y6H.m4s "
> 03.m4s
curl "https://video.twimg.com/dm_video/1600877027330064385/vid/9000/12000/320x180/V1cXuDCxjXKk_JT9.m4s
" > 04.m4s
curl "https://video.twimg.com/dm_video/1600877027330064385/vid/12000/15000/320x180/CtUTfpf83EHjEFjd.m4s
" > 05.m4s
curl "https://video.twimg.com/dm_video/1600877027330064385/vid/15000/18000/320x180/lGncur15MHC6fvKg.m4s
" > 06.m4s
curl "https://video.twimg.com/dm_video/1600877027330064385/vid/18000/21000/320x180/2U51PtuljYpAxhRr.m4s
" > 07.m4s
curl "https://video.twimg.com/dm_video/1600877027330064385/vid/21000/24000/320x180/gs-8pNvThX_1kjx0.m4s
" > 08.m4s
curl "https://video.twimg.com/dm_video/1600877027330064385/vid/24000/27000/320x180/f1UDV6NW3odTL-ux.m4s
" > 09.m4s
curl "https://video.twimg.com/dm_video/1600877027330064385/vid/27000/30000/320x180/9hFdPsP3QYV8c130.m4s
" > 10.m4s
curl "https://video.twimg.com/dm_video/1600877027330064385/vid/30000/32100/320x180/yFOkibojJs9PWhkX.m4s
" > 11.m4s
% ./run-me.sh
curl https://video.twimg.com/dm_video/1600877027330064385/vid/0/0/320x180/jZYOJeLERXPOC4qe.mp4
  % Total    % Received % Xferd  Average Speed   Time    Time     Time  Current
                                 Dload  Upload   Total   Spent    Left  Speed
100  1130  100  1130    0     0   6208      0 --:--:-- --:--:-- --:--:--  6208
curl https://video.twimg.com/dm_video/1600877027330064385/vid/0/3000/320x180/l1mZtezfzjRRYziE.m4s
  % Total    % Received % Xferd  Average Speed   Time    Time     Time  Current
                                 Dload  Upload   Total   Spent    Left  Speed
100 37919  100 37919    0     0   251k      0 --:--:-- --:--:-- --:--:--  251k
curl https://video.twimg.com/dm_video/1600877027330064385/vid/3000/6000/320x180/yxhsDmzuJG9ZqtYb.m4s
  % Total    % Received % Xferd  Average Speed   Time    Time     Time  Current
                                 Dload  Upload   Total   Spent    Left  Speed
100 35423  100 35423    0     0   283k      0 --:--:-- --:--:-- --:--:--  283k
curl https://video.twimg.com/dm_video/1600877027330064385/vid/6000/9000/320x180/KGtEzr2KaRfP4Y6H.m4s
  % Total    % Received % Xferd  Average Speed   Time    Time     Time  Current
                                 Dload  Upload   Total   Spent    Left  Speed
100 36960  100 36960    0     0   243k      0 --:--:-- --:--:-- --:--:--  243k
curl https://video.twimg.com/dm_video/1600877027330064385/vid/9000/12000/320x180/V1cXuDCxjXKk_JT9.m4s
  % Total    % Received % Xferd  Average Speed   Time    Time     Time  Current
                                 Dload  Upload   Total   Spent    Left  Speed
100 43395  100 43395    0     0   181k      0 --:--:-- --:--:-- --:--:--  181k
curl https://video.twimg.com/dm_video/1600877027330064385/vid/12000/15000/320x180/CtUTfpf83EHjEFjd.m4s
  % Total    % Received % Xferd  Average Speed   Time    Time     Time  Current
                                 Dload  Upload   Total   Spent    Left  Speed
100 47333  100 47333    0     0   185k      0 --:--:-- --:--:-- --:--:--  184k
curl https://video.twimg.com/dm_video/1600877027330064385/vid/15000/18000/320x180/lGncur15MHC6fvKg.m4s
  % Total    % Received % Xferd  Average Speed   Time    Time     Time  Current
                                 Dload  Upload   Total   Spent    Left  Speed
100 41711  100 41711    0     0   162k      0 --:--:-- --:--:-- --:--:--  162k
curl https://video.twimg.com/dm_video/1600877027330064385/vid/18000/21000/320x180/2U51PtuljYpAxhRr.m4s
  % Total    % Received % Xferd  Average Speed   Time    Time     Time  Current
                                 Dload  Upload   Total   Spent    Left  Speed
100 38884  100 38884    0     0   132k      0 --:--:-- --:--:-- --:--:--  132k
curl https://video.twimg.com/dm_video/1600877027330064385/vid/21000/24000/320x180/gs-8pNvThX_1kjx0.m4s
  % Total    % Received % Xferd  Average Speed   Time    Time     Time  Current
                                 Dload  Upload   Total   Spent    Left  Speed
100 36449  100 36449    0     0   144k      0 --:--:-- --:--:-- --:--:--  144k
```

```
curl https://video.twimg.com/dm_video/1600877027330064385/vid/24000/27000/320x180/f1UDV6NW3odTL-ux.m4s
  % Total    % Received % Xferd  Average Speed   Time    Time     Time  Current
                                 Dload  Upload   Total   Spent    Left  Speed
100 32279  100 32279    0     0  92755      0 --:--:-- --:--:-- --:--:-- 92755
curl https://video.twimg.com/dm_video/1600877027330064385/vid/27000/30000/320x180/9hFdPsP3QYV8c130.m4s
  % Total    % Received % Xferd  Average Speed   Time    Time     Time  Current
                                 Dload  Upload   Total   Spent    Left  Speed
100 32413  100 32413    0     0   109k      0 --:--:-- --:--:-- --:--:--  109k
curl https://video.twimg.com/dm_video/1600877027330064385/vid/30000/32100/320x180/yFOkibojJs9PWhkX.m4s
  % Total    % Received % Xferd  Average Speed   Time    Time     Time  Current
                                 Dload  Upload   Total   Spent    Left  Speed
100 22617  100 22617    0     0  94237      0 --:--:-- --:--:-- --:--:-- 93846
% cat 00.mp4 *.m4s > wk-bk-dm-video.mp4
% ls -l wk-bk-dm-video.mp4
-rw-r--r--@ 1 mln2  staff  406513 Dec  8 11:32 wk-bk-dm-video.mp4
% open wk-bk-dm-video.mp4
```

We also manually pushed the `.mp4` and `.m4s` files into the Wayback Machine using Save Page Now [2]. Note that it is also possible to submit large batches of URLs to be archived using Google Sheets [5]. Appendix 2 shows the result of retrieving the `.mp4` and `.m4s` files from the Wayback Machine and demonstrating that the resulting file is equivalent to the one retrieved from the live web.

We discovered that the Wayback Machine has video files (mostly `.mp4` but some `.m4s`) from approximately 100 unique DM ids, excluding the test id (1600877027330064385) we introduced for this study. The first known archived DM video is from 2016-03-04; it is possible there are earlier, undiscovered DM videos archived if the URL structure was different prior to March, 2016. Appendix 4 gives a summary of the archived DM videos.

# video.twimg.com Does Not Enforce HTTPS

Since the URLs of videos in DM are presumably hard to predict, guess, or reverse engineer, one of our concerns was how easy it would be to combine with another browser vulnerability, such as malicious extensions (e.g., [6, 7, 8]), to exfiltrate the video URLs. For example, changing the scheme of the video URLs from "https" to "http" would expose the URLs to eavesdroppers. In our investigations, we discovered that the host video.twimg.com does not require https authentication, and will serve content over http. This is not true for Twitter's main host, twitter.com. Neither twimg.com[2] nor video.twimg.com[3] are on the HSTS preload list. A Strict-Transport-Security response header is returned for requests to video.twimg.com:

```
% curl -I https://video.twimg.com/dm_video/1600877027330064385/vid/0/0/320x180/jZYOJeLERXPOC4qe.mp4
HTTP/2 200
perf: 7626143928
content-type: video/mp4
cache-control: max-age=604800, must-revalidate
last-modified: Thu, 08 Dec 2022 15:34:40 GMT
x-transaction-id: 95716829e6755cc1
```

---

[2] https://web.archive.org/web/20221222204832/https://hstspreload.org/?domain=twimg.com. Furthermore, at the time of this writing, neither www.twimg.com nor twimg.com are valid hostnames.
[3] https://web.archive.org/web/20221222203805/https://hstspreload.org/?domain=video.twimg.com

```
timing-allow-origin: https://twitter.com, https://mobile.twitter.com
x-content-type-options: nosniff
strict-transport-security: max-age=631138519
access-control-allow-origin: *
access-control-expose-headers: Content-Length
accept-ranges: bytes
date: Thu, 22 Dec 2022 20:58:32 GMT
x-served-by: cache-fty21371-FTY, cache-iad-kjyo7100080-IAD
x-cache: MISS, HIT
x-tw-cdn: FT
server-timing: x-cache;desc=HIT, x-tw-cdn;desc=FT
content-length: 1130
```

But video.twimg.com will still serve the content via http:

```
% curl -I http://video.twimg.com/dm_video/1600877027330064385/vid/0/3000/320x180/l1mZtezfzjRRYziE.m4s
HTTP/1.1 200 OK
Connection: keep-alive
Content-Length: 37919
perf: 7626143928
content-type: video/mp4
cache-control: max-age=604800, must-revalidate
last-modified: Thu, 08 Dec 2022 15:34:40 GMT
x-transaction-id: 4e772788c1fdbf85
timing-allow-origin: https://twitter.com, https://mobile.twitter.com
x-content-type-options: nosniff
strict-transport-security: max-age=631138519
access-control-allow-origin: *
access-control-expose-headers: Content-Length
Accept-Ranges: bytes
Date: Thu, 22 Dec 2022 20:59:53 GMT
X-Served-By: cache-fty21368-FTY, cache-iad-kjyo7100091-IAD
X-Cache: MISS, MISS
x-tw-cdn: FT
Server-Timing: x-cache;desc=MISS, x-tw-cdn;desc=FT

% curl -s http://video.twimg.com/dm_video/1600877027330064385/vid/0/3000/320x180/l1mZtezfzjRRYziE.m4s |
wc -c
   37919
```

However, it would not be possible for a malicious browser extension to make use of the still active http server since the connect-src attribute of the Content-Security-Policy response header from twitter.com restricts fetch() requests to https://*.twimg.com/ (bottom right, Figure 5).

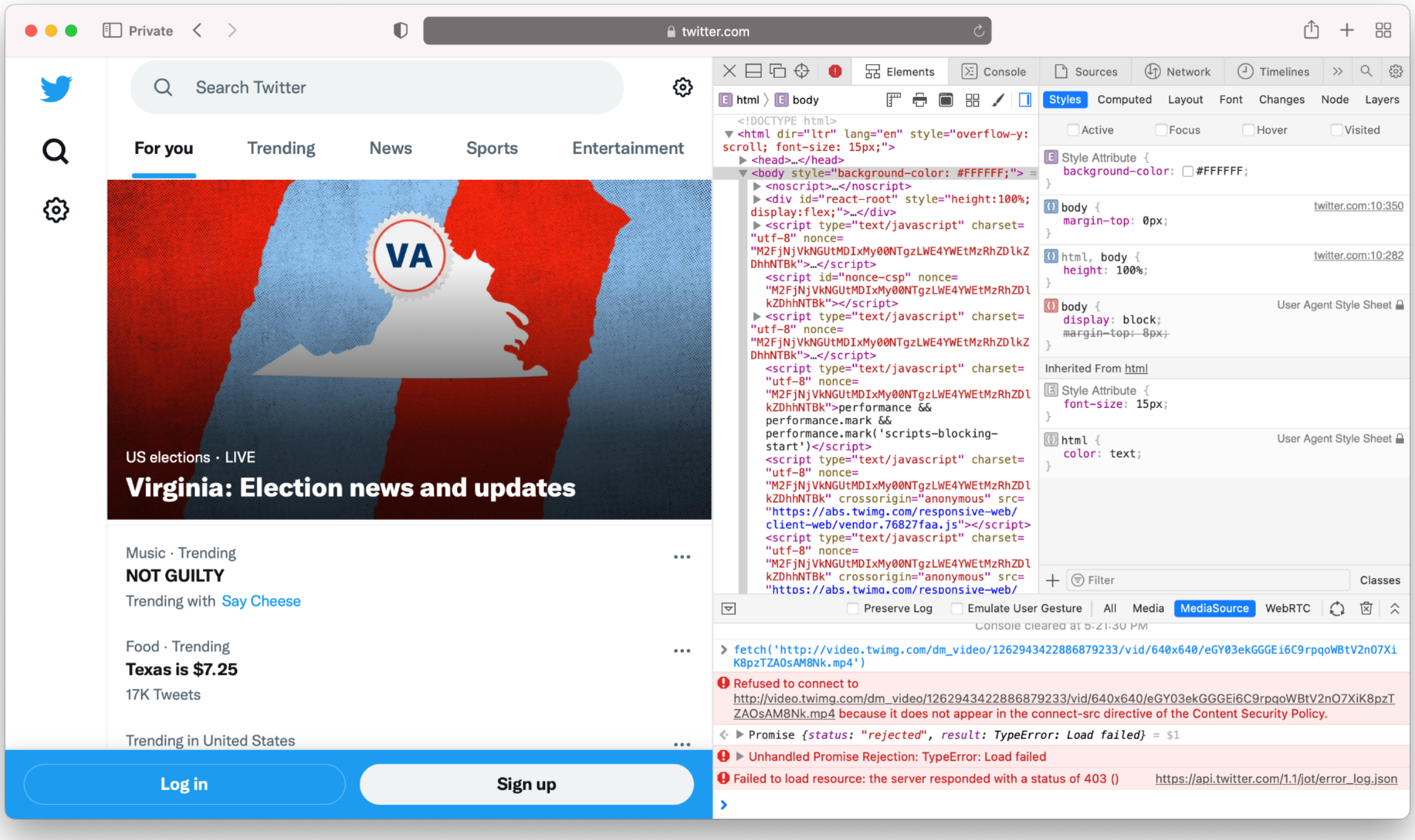

Figure 5. A fetch() request to [http://video.twimg.com/dm_video/1262943422886879233/vid/640x640/eGY03ekGGGEi6C9rpqoWBtV2nO7XiK8pzTZAOsAM8Nk.mp4](http://video.twimg.com/dm_video/1262943422886879233/vid/640x640/eGY03ekGGGEi6C9rpqoWBtV2nO7XiK8pzTZAOsAM8Nk.mp4) (note: http, not https) is blocked by twitter.com's Content-Security-Policy header ("Refused to connect…" in the bottom right).

# Future Work

One area of possible future exploration is deleting the DMs and/or the accounts (or creating and deleting similar accounts) and determining if the videos persist after the DMs and/or accounts are deleted (cf. [1]).

We could also explore if it is possible to uncover direct links to the unfragmented `.mp4` file from the developer tools and without an API call (which would likely be harder to exfiltrate). There are a number of services that do this for public videos (e.g., https://twittervideodownloader.com, https://ssstwitter.com, https://www.savetweetvid.com), but we are not aware of such services for videos in DMs.

We could also explore for how long image URLs will return an HTTP response after the session is terminated (or perhaps this is documented in the Twitter API).

# Conclusions

Twitter goes to great lengths to protect the images shared in DMs: they are only available to parties in the particular DM chat where they were shared, and only while the session is active (i.e., with at least one member viewing the session in their browser). This is accomplished by requiring a twitter.com HTTP Cookie for access to the images: twitter.com will return a 401 is the cookie is absent, a 404 if the cookie is present but not valid for that session, and return no HTTP response after some time has passed since the termination of the HTTP session.

Despite having similarly opaque naming conventions for URLs for both DM images and DM videos, the videos shared in DMs do not have the same level of HTTP protection. The URLs for videos shared in DMs are available without HTTP Cookies, and can be easily accessed if their URLs are somehow known. This behavior can be replicated by interested users by copying and running the scripts "`run-me.sh`" and/or "`run-me-ia.sh`".

We have created a minimal reproducible example using three demo accounts, where in a DM session two accounts have shared an image and a video. The image has HTTP protection, but the video does not. The video is downloaded with simple curl commands (without HTTP authentication or Cookies), as well as the URLs pushed to the Internet Archive's Wayback Machine. The video can be recovered and reassembled from both twitter.com and archive.org.

# Appendix 1

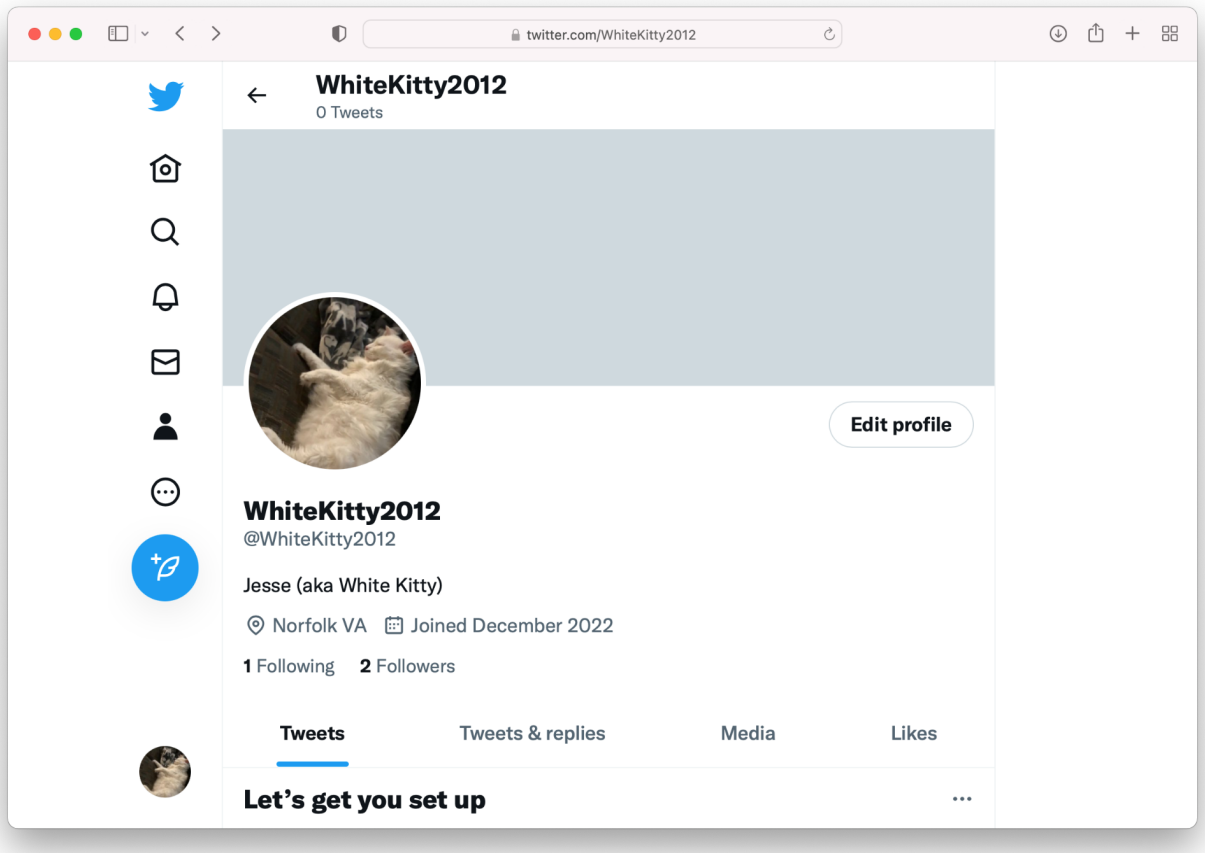

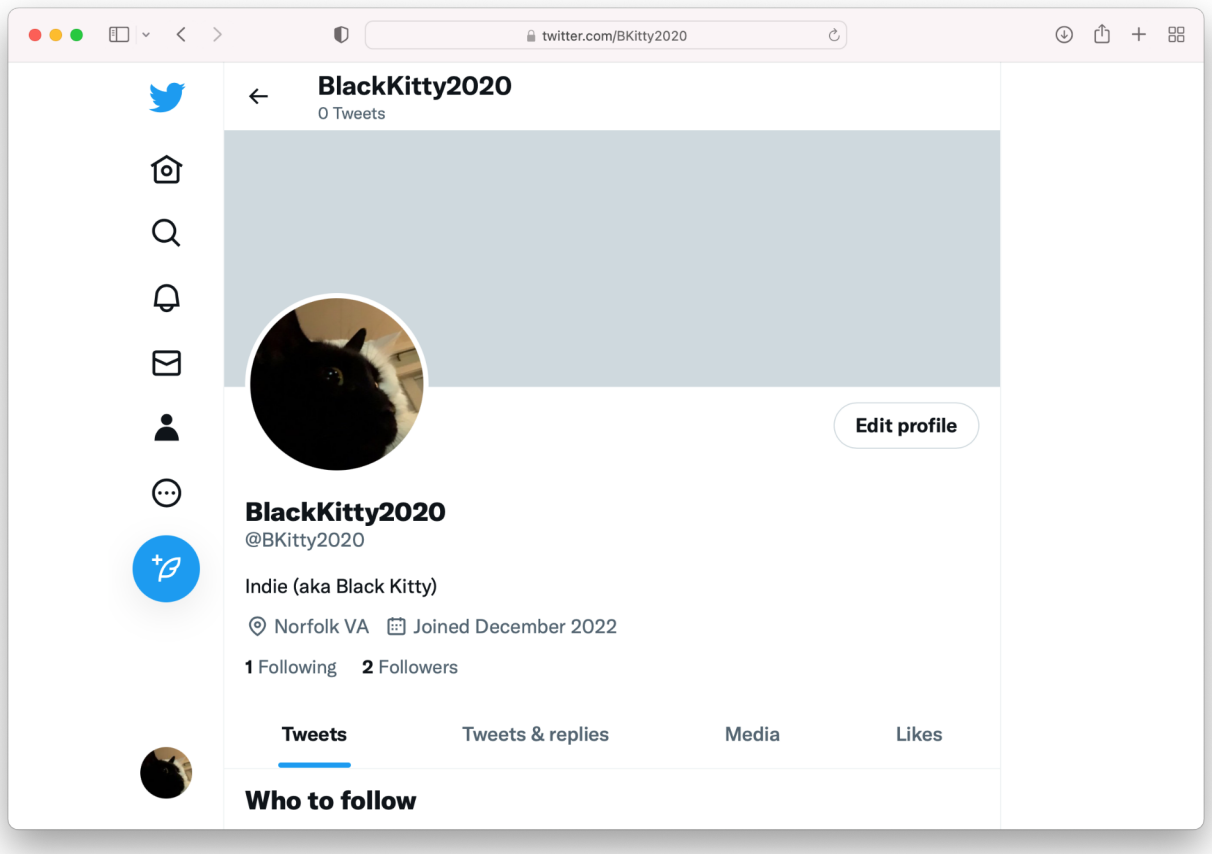

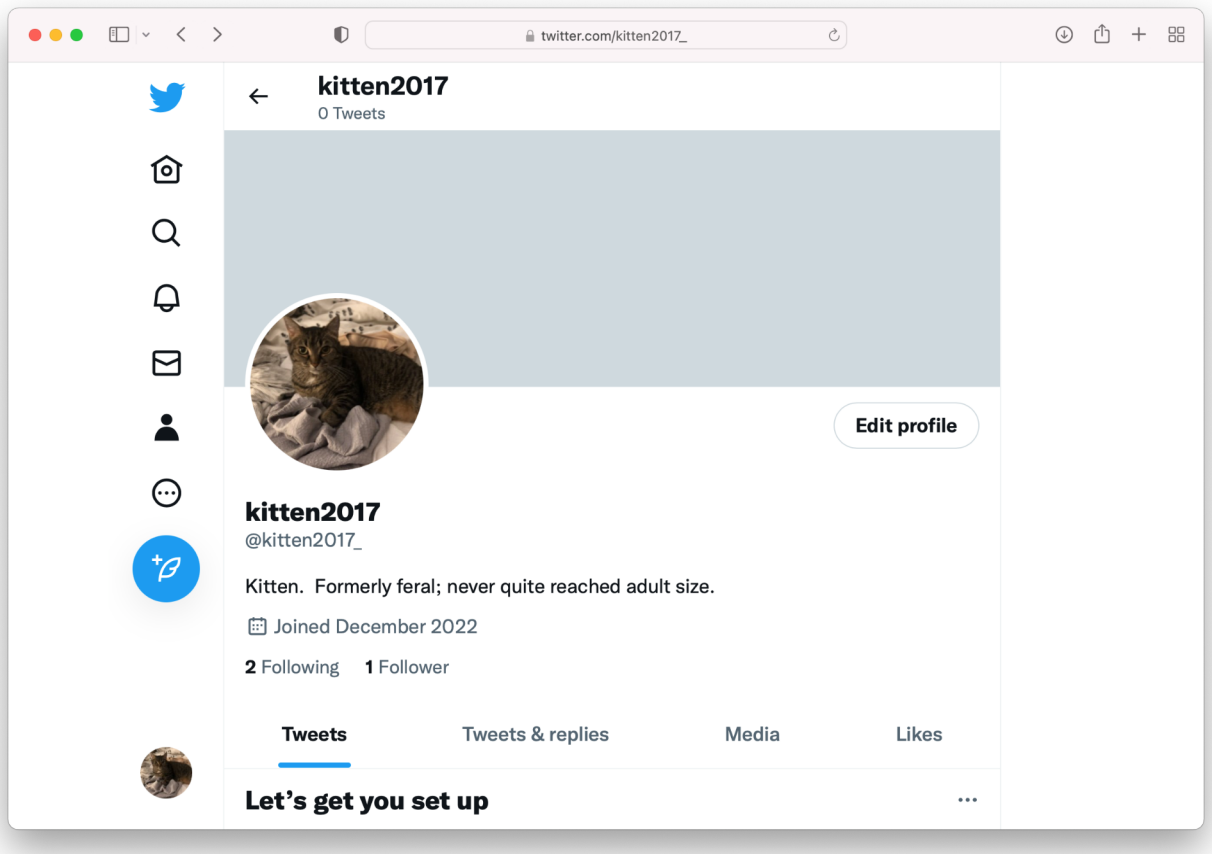

Twitter account pages for: @WhiteKitty2012, @BKitty2020, and @kitten2017_.

# Appendix 2

Retrieving copies of the `.mp4` and `.m4s` files from the Internet Archive Wayback Machine, concatenating them into a single `.mp4`, and showing that the resulting `.mp4` from archive.org is the same as the `.mp4` available from the live web (twitter.com).

```
% cat run-me-ia.sh
#!/bin/csh -x

curl -L
"https://web.archive.org/web/20221208194342id_/https://video.twimg.com/dm_video/1600877027330064385/vid
/0/0/320x180/jZYOJeLERXPOC4qe.mp4" > ia-00.mp4
curl -L
"https://web.archive.org/web/20221208194342id_/https://video.twimg.com/dm_video/1600877027330064385/vid
/0/3000/320x180/l1mZtezfzjRRYziE.m4s " > ia-01.m4s
curl -L
"https://web.archive.org/web/20221208194342id_/https://video.twimg.com/dm_video/1600877027330064385/vid
/3000/6000/320x180/yxhsDmzuJG9ZqtYb.m4s " > ia-02.m4s
curl -L
"https://web.archive.org/web/20221208194342id_/https://video.twimg.com/dm_video/1600877027330064385/vid
/6000/9000/320x180/KGtEzr2KaRfP4Y6H.m4s " > ia-03.m4s
curl -L
"https://web.archive.org/web/20221208194342id_/https://video.twimg.com/dm_video/1600877027330064385/vid
/9000/12000/320x180/V1cXuDCxjXKk_JT9.m4s " > ia-04.m4s
curl -L
"https://web.archive.org/web/20221208194342id_/https://video.twimg.com/dm_video/1600877027330064385/vid
/12000/15000/320x180/CtUTfpf83EHjEFjd.m4s " > ia-05.m4s
curl -L
"https://web.archive.org/web/20221208194342id_/https://video.twimg.com/dm_video/1600877027330064385/vid
/15000/18000/320x180/lGncur15MHC6fvKg.m4s " > ia-06.m4s
curl -L
"https://web.archive.org/web/20221208194342id_/https://video.twimg.com/dm_video/1600877027330064385/vid
/18000/21000/320x180/2U51PtuljYpAxhRr.m4s " > ia-07.m4s
curl -L
"https://web.archive.org/web/20221208194342id_/https://video.twimg.com/dm_video/1600877027330064385/vid
/21000/24000/320x180/gs-8pNvThX_1kjx0.m4s " > ia-08.m4s
curl -L
"https://web.archive.org/web/20221208194342id_/https://video.twimg.com/dm_video/1600877027330064385/vid
/24000/27000/320x180/f1UDV6NW3odTL-ux.m4s " > ia-09.m4s
curl -L
"https://web.archive.org/web/20221208194342id_/https://video.twimg.com/dm_video/1600877027330064385/vid
/27000/30000/320x180/9hFdPsP3QYV8c130.m4s " > ia-10.m4s
curl -L
"https://web.archive.org/web/20221208194342id_/https://video.twimg.com/dm_video/1600877027330064385/vid
/30000/32100/320x180/yFOkibojJs9PWhkX.m4s " > ia-11.m4s
% ./run-me-ia.sh
curl -L
https://web.archive.org/web/20221208194342id_/https://video.twimg.com/dm_video/1600877027330064385/vid/
0/0/320x180/jZYOJeLERXPOC4qe.mp4
  % Total    % Received % Xferd  Average Speed   Time    Time     Time  Current
                                 Dload  Upload   Total   Spent    Left  Speed
  0     0    0     0    0     0      0      0 --:--:--  0:00:03 --:--:--     0
100  1130  100  1130    0     0    255      0  0:00:04  0:00:04 --:--:--  3434
curl -L
https://web.archive.org/web/20221208194342id_/https://video.twimg.com/dm_video/1600877027330064385/vid/
0/3000/320x180/l1mZtezfzjRRYziE.m4s
```

```
  % Total    % Received % Xferd  Average Speed   Time    Time     Time  Current
                                 Dload  Upload   Total   Spent    Left  Speed
  0     0    0     0    0     0      0      0 --:--:-- --:--:-- --:--:--     0
100 37919  100 37919    0     0  18143      0  0:00:02  0:00:02 --:--:-- 77071
curl -L
https://web.archive.org/web/20221208194342id_/https://video.twimg.com/dm_video/1600877027330064385/vid/
3000/6000/320x180/yxhsDmzuJG9ZqtYb.m4s
  % Total    % Received % Xferd  Average Speed   Time    Time     Time  Current
                                 Dload  Upload   Total   Spent    Left  Speed
  0     0    0     0    0     0      0      0 --:--:-- --:--:-- --:--:--     0
100 35423  100 35423    0     0  17355      0  0:00:02  0:00:02 --:--:-- 96520
curl -L
https://web.archive.org/web/20221208194342id_/https://video.twimg.com/dm_video/1600877027330064385/vid/
6000/9000/320x180/KGtEzr2KaRfP4Y6H.m4s
  % Total    % Received % Xferd  Average Speed   Time    Time     Time  Current
                                 Dload  Upload   Total   Spent    Left  Speed
  0     0    0     0    0     0      0      0 --:--:--  0:00:01 --:--:--     0
100 36960  100 36960    0     0  16332      0  0:00:02  0:00:02 --:--:-- 16332
curl -L
https://web.archive.org/web/20221208194342id_/https://video.twimg.com/dm_video/1600877027330064385/vid/
9000/12000/320x180/V1cXuDCxjXKk_JT9.m4s
  % Total    % Received % Xferd  Average Speed   Time    Time     Time  Current
                                 Dload  Upload   Total   Spent    Left  Speed
  0     0    0     0    0     0      0      0 --:--:--  0:00:01 --:--:--     0
100 43395  100 43395    0     0  24094      0  0:00:01  0:00:01 --:--:-- 24094
curl -L
https://web.archive.org/web/20221208194342id_/https://video.twimg.com/dm_video/1600877027330064385/vid/
12000/15000/320x180/CtUTfpf83EHjEFjd.m4s
  % Total    % Received % Xferd  Average Speed   Time    Time     Time  Current
                                 Dload  Upload   Total   Spent    Left  Speed
  0     0    0     0    0     0      0      0 --:--:-- --:--:-- --:--:--     0
100 47333  100 47333    0     0  34676      0  0:00:01  0:00:01 --:--:--  276k
curl -L
https://web.archive.org/web/20221208194342id_/https://video.twimg.com/dm_video/1600877027330064385/vid/
15000/18000/320x180/lGncur15MHC6fvKg.m4s
  % Total    % Received % Xferd  Average Speed   Time    Time     Time  Current
                                 Dload  Upload   Total   Spent    Left  Speed
  0     0    0     0    0     0      0      0 --:--:--  0:00:02 --:--:--     0
100 41711  100 41711    0     0  14407      0  0:00:02  0:00:02 --:--:--     0
curl -L
https://web.archive.org/web/20221208194342id_/https://video.twimg.com/dm_video/1600877027330064385/vid/
18000/21000/320x180/2U51PtuljYpAxhRr.m4s
  % Total    % Received % Xferd  Average Speed   Time    Time     Time  Current
                                 Dload  Upload   Total   Spent    Left  Speed
  0     0    0     0    0     0      0      0 --:--:-- --:--:-- --:--:--     0
100 38884  100 38884    0     0  24957      0  0:00:01  0:00:01 --:--:--  303k
curl -L
https://web.archive.org/web/20221208194342id_/https://video.twimg.com/dm_video/1600877027330064385/vid/
21000/24000/320x180/gs-8pNvThX_1kjx0.m4s
  % Total    % Received % Xferd  Average Speed   Time    Time     Time  Current
                                 Dload  Upload   Total   Spent    Left  Speed
  0     0    0     0    0     0      0      0 --:--:--  0:00:01 --:--:--     0
100 36449  100 36449    0     0  19596      0  0:00:01  0:00:01 --:--:--  348k
curl -L
https://web.archive.org/web/20221208194342id_/https://video.twimg.com/dm_video/1600877027330064385/vid/
24000/27000/320x180/f1UDV6NW3odTL-ux.m4s
  % Total    % Received % Xferd  Average Speed   Time    Time     Time  Current
                                 Dload  Upload   Total   Spent    Left  Speed
  0     0    0     0    0     0      0      0 --:--:--  0:00:02 --:--:--     0
100 32279  100 32279    0     0   8039      0  0:00:04  0:00:04 --:--:-- 91962
curl -L
https://web.archive.org/web/20221208194342id_/https://video.twimg.com/dm_video/1600877027330064385/vid/
27000/30000/320x180/9hFdPsP3QYV8c130.m4s
  % Total    % Received % Xferd  Average Speed   Time    Time     Time  Current
                                 Dload  Upload   Total   Spent    Left  Speed
  0     0    0     0    0     0      0      0 --:--:--  0:00:01 --:--:--     0
```

```
100 32413  100 32413    0     0   9866      0  0:00:03  0:00:03 --:--:-- 45080
curl -L
https://web.archive.org/web/20221208194342id_/https://video.twimg.com/dm_video/1600877027330064385/vid/
30000/32100/320x180/yFOkibojJs9PWhkX.m4s
  % Total    % Received % Xferd  Average Speed   Time    Time     Time  Current
                                 Dload  Upload   Total   Spent    Left  Speed
100 22617  100 22617    0     0  13790      0  0:00:01  0:00:01 --:--:-- 13782
% ls -l ia-??.*
-rw-r--r--  1 mln2  staff   1130 Dec  8 14:51 ia-00.mp4
-rw-r--r--  1 mln2  staff  37919 Dec  8 14:52 ia-01.m4s
-rw-r--r--  1 mln2  staff  35423 Dec  8 14:52 ia-02.m4s
-rw-r--r--  1 mln2  staff  36960 Dec  8 14:52 ia-03.m4s
-rw-r--r--  1 mln2  staff  43395 Dec  8 14:52 ia-04.m4s
-rw-r--r--  1 mln2  staff  47333 Dec  8 14:52 ia-05.m4s
-rw-r--r--  1 mln2  staff  41711 Dec  8 14:52 ia-06.m4s
-rw-r--r--  1 mln2  staff  38884 Dec  8 14:52 ia-07.m4s
-rw-r--r--  1 mln2  staff  36449 Dec  8 14:52 ia-08.m4s
-rw-r--r--  1 mln2  staff  32279 Dec  8 14:52 ia-09.m4s
-rw-r--r--  1 mln2  staff  32413 Dec  8 14:52 ia-10.m4s
-rw-r--r--  1 mln2  staff  22617 Dec  8 14:52 ia-11.m4s
% cat ia-00.mp4 ia-*m4s > ia-wk-bk-dm-video.mp4
% ls -l wk-bk-dm-video.mp4 ia-wk-bk-dm-video.mp4
-rw-r--r--  1 mln2  staff  406513 Dec  8 14:52 ia-wk-bk-dm-video.mp4
-rw-r--r--@ 1 mln2  staff  406513 Dec  8 11:32 wk-bk-dm-video.mp4
% diff -s wk-bk-dm-video.mp4 ia-wk-bk-dm-video.mp4
Files wk-bk-dm-video.mp4 and ia-wk-bk-dm-video.mp4 are identical
```

# Appendix 3

This appendix contains the response we received to "#1798935 Twitter DM Videos Are Accessible to Unauthenticated Users", submitted at HackerOne.[4] We submitted our paper to arXiv.org after receiving this response.

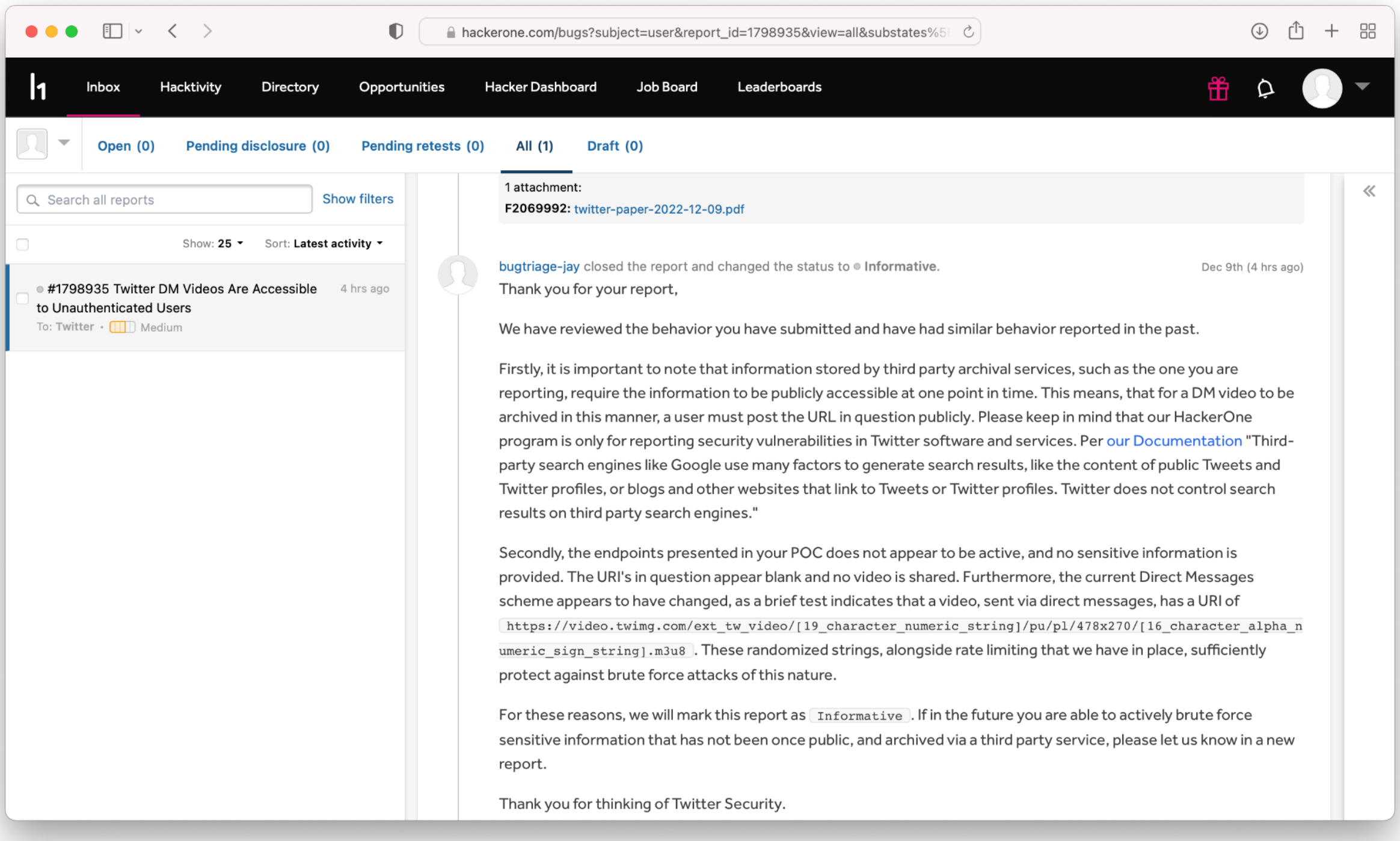

hackerone.com/bugs?subject=user&report_id=1798935&view=all&substates%5

Inbox | Hacktivity | Directory | Opportunities | Hacker Dashboard | Job Board | Leaderboards

Open (0) | Pending disclosure (0) | Pending retests (0) | All (1) | Draft (0)

Search all reports | Show filters

Show: 25 | Sort: Latest activity

#1798935 Twitter DM Videos Are Accessible to Unauthenticated Users — 4 hrs ago
To: Twitter • Medium

1 attachment:
F2069992: twitter-paper-2022-12-09.pdf

bugtriage-jay closed the report and changed the status to Informative. — Dec 9th (4 hrs ago)

Thank you for your report,

We have reviewed the behavior you have submitted and have had similar behavior reported in the past.

Firstly, it is important to note that information stored by third party archival services, such as the one you are reporting, require the information to be publicly accessible at one point in time. This means, that for a DM video to be archived in this manner, a user must post the URL in question publicly. Please keep in mind that our HackerOne program is only for reporting security vulnerabilities in Twitter software and services. Per our Documentation "Third-party search engines like Google use many factors to generate search results, like the content of public Tweets and Twitter profiles, or blogs and other websites that link to Tweets or Twitter profiles. Twitter does not control search results on third party search engines."

Secondly, the endpoints presented in your POC does not appear to be active, and no sensitive information is provided. The URI's in question appear blank and no video is shared. Furthermore, the current Direct Messages scheme appears to have changed, as a brief test indicates that a video, sent via direct messages, has a URI of `https://video.twimg.com/ext_tw_video/[19_character_numeric_string]/pu/pl/478x270/[16_character_alpha_numeric_sign_string].m3u8`. These randomized strings, alongside rate limiting that we have in place, sufficiently protect against brute force attacks of this nature.

For these reasons, we will mark this report as `Informative`. If in the future you are able to actively brute force sensitive information that has not been once public, and archived via a third party service, please let us know in a new report.

Thank you for thinking of Twitter Security.

[4] https://hackerone.com/reports/1798935

# Appendix 4

This appendix shows the commands used to list the DM videos already archived in the Internet Archive's Wayback Machine. We do not show the full list of URLs, since it is possible that not all of the original parties to these videos in Twitter DMs intended for the videos to be archived. It is also possible that some of these videos were also publicly shared videos, but were archived under their DM video URLs. DM videos have the prefix:

```
video.twimg.com/dm_video/
```

and public Twitter videos have the prefix:

```
video.twimg.com/ext_tw_video/
```

The command below accesses the Internet Archive's CDX API, does a "prefix" search on the DM video pattern, the awk command converts the CDX entry to web.archive.org URLs, the grep command blocks the DM id we introduced for this study, the sort command numerically sorts the DM ids and "uniques" them, and the wc command counts how many lines are left.

```
% curl -s
"http://web.archive.org/cdx/search/cdx?url=video.twimg.com/dm_video/&matchType=prefix" | awk
'{print "https://web.archive.org/web/" $2 "/" $3};' | grep -v "1600877027330064385" | sort -n
-k 10 -t "/" -u | wc -l
     103
```

The following command shows the first two lines (with the second line partially redacted):

```
% curl -s "http://web.archive.org/cdx/search/cdx?url=video.twimg.com/dm_video/&matchType=prefix" | awk
'{print "https://web.archive.org/web/" $2 "/" $3};' | grep -v "1600877027330064385" | sort -n -k 10 -t
"/" -u | head -2
https://web.archive.org/web/20201231013750/http://video.twimg.com/dm_video/13443
https://web.archive.org/web/20160304122159/https://video.twimg.com/dm_video/70284737[redacted].mp4?_=1
```

The first URL is incomplete and is an error (*shown in italics*), and dereferencing that URL will return an archived 404 response. The second line shows that the first known archived DM video occurred on 2016-03-04. Our cursory, manual inspection revealed that many, but not all, of these archived videos are still available on the live web (e.g., from video.twimg.com).

We are unsure of when videos in Twitter DMs were first supported, or if DM videos have had other URL prefixes.